\documentclass[conference]{IEEEtran}
\usepackage{times,amsmath,epsfig,latexsym,amssymb,psfrag,graphicx}
\usepackage{paralist}
\usepackage{booktabs}
\usepackage{color}
 \usepackage{setspace}
\usepackage{epstopdf}
\usepackage{subcaption}
\usepackage{mathbbol}
\usepackage[ruled,vlined]{algorithm2e}
\usepackage{citesort}
\newcommand{\qed}{\nobreak \ifvmode \relax \else
\ifdim\lastskip<1.5em \hskip-\lastskip
\hskip1.5em plus0em minus0.5em \fi \nobreak
\vrule height0.75em width0.5em depth0.25em\fi}
\begin{document}

\newcommand{\CS}[1]{\textcolor{magenta}{#1}} 
\newcommand{ \PP}[1]{\textcolor{blue}{#1}} 
\newcommand{\AZ}[1]{\textcolor{green}{#1}} 

\title{User Traffic Prediction for Proactive   Resource Management:  Learning-Powered Approaches}
\author{Amin Azari$^*$, Panagiotis Papapetrou$^*$, \v Stojan Denic$^+$, and Gunnar Peters$^+$\\
$^*$Stockholm University, Sweden $^+$Huawei, Sweden \\
Email: \{amin.azari,Panagiotis\}@dsv.su.se, \{stojan.denic,gunnar.peters\}@huawei.com}
\maketitle
 
\maketitle           
\begin{abstract}
Traffic prediction  plays a vital role in efficient planning and usage of  network  resources in wireless networks. While traffic prediction in wired networks is an established field, there is a lack of research on the analysis of traffic in cellular networks, especially in a content-blind manner at  the user level. Here, we shed light into this problem by designing  traffic prediction  tools that employ either  statistical, rule-based, or deep machine learning  methods. First, we present an extensive experimental evaluation of the designed tools over a real traffic dataset. Within this analysis, the impact of different parameters, such as  length of prediction, feature set used in analyses, and granularity of data, on accuracy of prediction   are investigated. Second, regarding the coupling observed between behavior of traffic and its generating application, we extend our analysis to the blind classification of  applications generating the traffic  based on the statistics of traffic arrival/departure. The results demonstrate presence of a threshold number of previous observations, beyond which, deep machine learning can outperform linear statistical  learning, and before which, statistical learning outperforms deep learning  approaches. Further analysis of this threshold value represents a strong coupling between this threshold, the length of future prediction, and the feature set in use. Finally, through a case study, we present how the experienced delay could be decreased by traffic arrival prediction.

\end{abstract}

\section{Introduction}\label{sec:introduction}

A major driver for the fifth generation (5G) of wireless networks and beyond consists in offering a wide set of cellular services in an energy and cost efficient way \cite{Sad6g}. Toward this end, the legacy design approach, in which resource provisioning and operation control are performed based on the peak traffic scenarios, are substituted with predictive analysis of mobile network traffic and proactive network resource management \cite{RaUrllc,Sad6g}. Indeed, in cellular networks with limited and highly expensive time-frequency radio resources, precise prediction of user traffic arrival can contribute significantly in improving the radio resource utilization and moving towards  cognitive and autonomous wireless access networks  \cite{RaUrllc}. As a result, in recent years, there has been an increasing interest in leveraging machine learning tools in analyzing the aggregated traffic served in a service area for optimizing the operation of the network \cite{RnnScale,DrxSelf,DrlBsSleep,UavML}.  Scaling of fronthaul and backhaul resources for 5G networks has been investigated in \cite{RnnScale} by leveraging methods from recurrent neural networks (RNNs) for traffic estimation. Analysis of cellular traffic for finding  anomaly in the performance and provisioning of on-demand resources for compensating such anomalies have been investigated  in \cite{UavML}. Furthermore,  prediction of light-traffic periods, and saving energy for base stations (BSs) through sleeping them in the respective periods has been investigated in \cite{DrxSelf,DrlBsSleep}.  While one observes that analysis of the aggregated traffic at the network side is an established field, there is lack of research on the analysis and understanding at the user level, i.e., of the specific users' traffic arrival. In 5G-and-beyond networks, the (i) explosively growing demand for radio access,  (ii) intention for serving battery- and  radio-limited devices requiring low-cost energy efficient service  \cite{SelfMl},  and  (iii) intention for supporting ultra-reliable low-latency communications (URLLC) \cite{RaUrllc}, mandate studying not only the aggregated traffic arrival from users, but also studying the features of traffic arrival in each user, or at least for critical users. A critical user could be defined as a user whose quality-of-service (QoS) is at risk due to the traffic behavior of other devices, or its behavior affects the QoS of other users, which is usually the case in URLLC scenarios \cite{RaUrllc}. 

 The traffic analysis and prediction problem could be approached as a  time series forecasting problem, where for example, the number of packet arrivals in each unit of time could be defined as the value of the time series at each point. While the literature on time series forecasting using statistical and machine learning approaches is mature \cite{bookrima,HybArimaLstm}, understanding  dynamics of cellular traffic and prediction of  future traffic/burst arrivals are complex problems. This is mainly because of the  vastly diverse set of parameters that shape the traffic arrival process, from set of running applications in the background to the communication system in use. Dealing with cellular traffic prediction as a time series prediction, one may categorize the state-of-the-art proposed schemes into three categories: statistical learning \cite{bookrima}, machine learning \cite{LstmTrafficRaw,SpTmBigDL}, and hybrid schemes \cite{HybArimaLstm}. ARIMA and LSTM, as two well-known statistical and machine learning approaches, respectively, for forecasting time series, which have been compared in a variety of problems, from economics to network engineering  \cite{NeuTM}. A comprehensive survey on cellular traffic prediction schemes could be found in \cite{DlForecast,EntropyCellular}. A deep learning-powered approach for prediction of overall network demand in each region of cities has been proposed in \cite{NetDemand}. In \cite{SpTmTraffic,SpTmBigDL}, the spatial and temporal correlations of the cellular traffic in different time periods and neighbouring cells, respectively, have been explored using neural networks in order to improve the accuracy of traffic prediction.  In \cite{ModelLstmDnn}, convolutional and recurrent neural networks have been combined in order to further capture dynamics of time series, and enhance the prediction performance. In \cite{NeuTM,LstmTrafficRaw}, the aggregated network traffic prediction using LSTM have been presented, while the study on the feature sets used in the experiment and the impact of different design parameters on the performance are missing. Study of state-of-the-art reveals that  there is a lack of research on leveraging advanced learning tools for cellular traffic prediction, selection of adequate features, especially when it comes to each user with specific set applications, which is covered here. 

In this work, we present our preliminary results on generation, labeling, and analysis of cellular traffic captured from a real user using deep machine learning as well as statistical learning. The main contributions of this work include:
\begin{itemize}
\item
Formulate the  traffic analysis problem as a time series classification/forecasting problem, design a set of features based on traffic statistics, and leverage statistical and deep learning  for approaching this problem.
\item 
Generate a real  labeled  traffic dataset, carry out a comprehensive set of traffic  analysis, including: (i) performance comparison of  deep-learning predictor against linear statistical-learning predictor, in terms of short-term and long-term predictive performance; (ii) performance analysis of adding extra  features to the deep learning predictors; (iii) analysis of tuning design parameters, e.g. the \textit{length} of previous \textit{observations} and future \textit{prediction} on the prediction performance.
\item
Identify a threshold number of previous observations, before which, statistical learning outperform deep learning, and identify the coupling between this threshold and the \textit{length} of future prediction and the  feature set is use.
\item
Present the usefulness of traffic-aware radio resource management through a case study, and investigate how the experienced delay in communications could be decreased by predicting the arrival of bursts.  
\end{itemize}
The remainder of this paper is organized as follows. The next section presents the problem description and the structure of proposed solution. Section 3 presents the set of methods used in the solution. Section 4 presents the experimental  results. Finally, the concluding remarks  are given in Section 5.


\section{Problem description and  traffic prediction framework} 
\subsection{Problem description} \label{sec:prob}
In this section, we first  introduce the research problem addressed in the paper. Then, we present the  structure of the proposed solution for addressing this problem. The system model considered in this work consists of a set of wireless devices connected to a cellular network, on which, a set of applications are running. 
At a given time interval $[t,t+\tau]$ with length $\tau$, each application could be in an \emph{active} or \emph{inactive} mode, based on the user behavior, and  traffic generation of the application is dependent upon its activity mode. The aim is to control the level of available network resources, e.g. the amount of radio resources, and to allocate the available resources to the devices demanding service.  The legacy approaches for  network resource management usually provision and allocate resources based on the buffer status report (BSR) of users. However, here we put one step forward and seek for opportunities to carry proactive resource provisioning and allocation out.  Then, given the current and past state of users' traffic behavior, we aim at making decision for serving a coexistence set of users. Towards this end, we need to carry an in-depth analysis of individual user's traffic behavior. Let us denote the set of per-user features describing aggregated  cellular traffic in  $[t,t+\tau]$ by $\textbf x(t)$. Furthermore, let $\textbf{X}_m(t)$ denote a matrix containing the latest $m$ feature vectors of traffic for $m\ge 0$. For example, $\textbf{X}_2(t)=[\textbf{x}(t-1),\textbf{x}(t)]$. Further, denote by $\textbf s$ an indicator vector, with elements either 0 or 1. Then, given a matrix $\textbf{X}_m(t)$ and a binary indicator vector $\textbf s$, we define $\textbf{X}^s_m(t)$ the submatrix of $\textbf{X}_m(t)$, such that all respective rows, for which $s$ indicates a zero value, are removed.  For example, let
$
\textbf{X}_m(t)=[1,2;3,4]\text{     and      }
s=[1,0], $
then, $\textbf{X}^s_m(t)=[1,2]$.  
Now, the problem is formulated as:
\begin{align} 
\textrm{Given}\quad \textbf{X}_{m}(t\text{-}1); \textrm{ Minimize}\quad L\big({\textbf X}_{-n}^\textbf{s}(t),{\textbf Y(t)}\big)\label{opp},  
\end{align}
where $m\ge 0, n\in \mathbb{Z}, n\ge 0$ is the length of the future predictions, e.g., $m=0$ for one step prediction, $\textbf Y(t)$ is of the same size as ${\textbf X}_{-n}^\textbf{s}(t)$ and represents the predicted matrix at time $t$, while $L(\cdot)$ is the desired error function, e.g., it may compute the mean squared error  between ${\textbf X}_{-n}^\textbf{s}(t)$ and $\textbf Y(t)$. 
 \begin{figure}[!t]
        \centering
                \includegraphics[width=0.95\columnwidth]{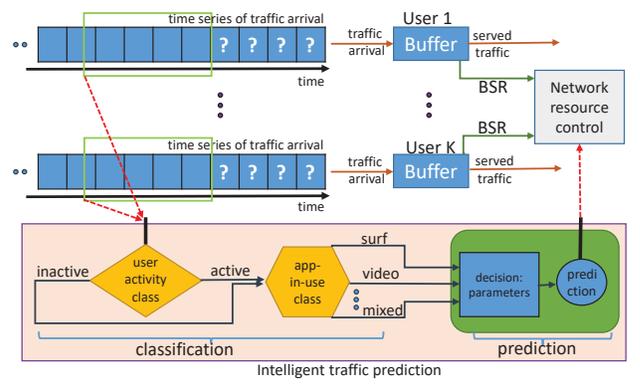}
                \caption{The proposed  intelligent module for enhanced  decision making in control of network resources for serving users } 
                \label{pr}
\end{figure} 

\subsection{The overall structure of the proposed solution} \label{sec:prop}
The main challenges, as described in the previous section, in the prediction of cellular traffic consists in dependency of traffic arrival to user behavior and type of the application(s) generating the traffic.  Then, as part of the solution to this problem, one may first predict the application(s) in use and behavior of the user, and then extract  extra features based on the classification of users' and applications mode, to be leveraged in the traffic prediction. This approach for solving \eqref{opp} has been  illustrated in Fig.~\ref{pr}. In order to realize such a framework, it is of crucial importance to first evaluate the traffic predictability and classficablity using only statistics of traffic, and then, to investigate hybrid models for augmenting predictors by online classifications, and finally to investigate  traffic-aware network management design. In the following sections, predictability and classficablity of cellular traffic  is investigated, a case study of  predictive traffic serving is presented, and the other parts of the proposed framework are left for the extended version of manuscript.

\section{Traffic Prediction-Classification: Methods} \label{sec:methods}
\subsection{Statistical learning: ARIMA} \label{sec:arima}
The first method we consider in our work is Autoregressive integrated moving average (ARIMA), which is essentially a statistical regression model.  The predictions performed by ARIMA are based on considering the lagged values of a given time series, while at the same time accommodating non-stationarity. An ARIMA model, $ARIMA(p, d, q)$, is defined by three parameters $p,d,q$ \cite{mills1991time}, where $p$ and $q$ correspond to the AR and MA processes, respectively, while $d$ is the number of differentiations  performed to the original time series values, that is $X_t$ is converted to $X_t^{(d)} =\nabla^d X_t$, with $X_t^{(d)}$ being the time series value at time $t$, with differentiation applied $d$ times.  Consequently, the full $ARIMA(p, d, q)$ model is computed as follows:
$
 X_t^{(d)} = \sum\nolimits_{i=1}^{p}\alpha_iX_{t-i}^{(d)}+\sum\nolimits_{j=1}^{q}\beta_j \epsilon_{t-j}+
 \epsilon_t + c+ \mu.$
In this  study, ARIMA is used for traffic prediction, and the ARIMA parameters, including $p$, $d$, and $q$, are optimized by carrying out a grid search over potential values in order to locate the best set of parameters.

\subsection{Rule-Based Learning: Decision trees and random forests} \label{sec:rf}
A decision tree is a rule-based classifier, where each internal node corresponds to a condition on
	a data attribute. The outcome of the condition can be binary, categorical (nominal or ordinal), or real-valued. Depending on the outcome of the condition
	the test example follows the corresponding branch, starting from the root node all the way down to a leaf node. Leaf nodes contain a class label, which correspond to the final classification outcome. A path from the root node to a leaf node builds a decision rule. The idea of a single decision tree is extended naturally to random forests (RAF)s and ensemble learning, based on 	the key fact that using an ensemble of many simple weak classifiers can lead to a much stronger classifier, given that each individual weak classifier is slightly stronger than random guessing and independent of all other classifiers. To classify a new object, it is sent to each tree in the forest, and each tree gives a result. The final class label is
determined by majority voting. More formally, let $h_i$ be a single learner, i.e. in our case a decision tree. Given a data example $x$, the RAF 
determines the final class label as follows using a set of $k$ independent decision trees, as follows: $R(x) = M^{\star} \{h_1 (x), \ldots, h_k(x)\},$ where $M^{\star}$ denotes the majority vote function of the set of individual learners. In this study, RAF is used for traffic classification.

\subsection{Deep learning: LSTM} \label{sec:lstm}
The second method we consider in our study is a long short-term memory (LSTM) architecture, which is based on a Recurrent Neural Network (RNN), a generalization of the feed forward network model for dealing with sequential data, with the addition of an ongoing internal state serving as a memory buffer for processing sequences. Let $\{X_1, \ldots, X_n\}$ define the set of $n$ time series inputs of our RNN and $\{Y_1, \ldots, Y_n\}$ be the set of outputs. For this study the internal state of the network is processed by Gated Recurrent Units (GRU)  defined by iterating the following three 
equations:
\begin{align}
&r_j \text{=}\text{S}([W_rX]_j \text{+} [U_rh_{t-1}]_j); z_j\text{=}\text{S}([W_zX]_j \text{+} [U_zh_{t-1}]_j));\nonumber\\
&h_j^t \text{=} z_jh_j^{t-1} \text{+} (1\text{-}z_j)h_\text{new}; h_\text{new}^t \text{=} \text{tanh}([WX]_j \text{+} [U(r \circ h_{t\text{-}1})]_j),\nonumber
\end{align}
where $r_j$ is a reset gate, $h_{t-1}$ is the previous hidden internal state $h_{t-1}$, $W$ and $U$  contain weights to be learned by the network, $z_j$ is an update gate, $h_j^t$ denotes the activation function of hidden unit $h_j$, $\text{S}(\cdot)$ denotes the sigmod function, and $\circ$ is the Hadamard product.
Finally, the loss function we optimize is the squared error, defined for all inputs as  $ \mathcal{L} = \sum\nolimits_{t=1}^n (Y_t-Y_t')^2 \ .$ In this study, LSTM is used for both traffic classification and prediction.

\section{Experimental evaluation} \label{sec:experiments}

\subsection{Traffic prediction and classification}
\subsubsection{Dataset, features, and feature sets} \label{sec:dataset}
For setting up any prediction tool, having access to a large and well-representative dataset is of crucial importance. Reviewing the state-of-the-art, as well as online resources, reveals that to the best of authors' knowledge, there are no public  datasets available representing cellular traffic to/from a user. Among several other reasons, privacy is a major reason that results in a lack of availability of cellular traffic records of  users. Then, in order to carry this research out, in this work we generate our own dataset and made it available online \cite{GHAA} for future works.  In order to generate the dataset, we leverage a packet capture tool, e.g. WireShark, at the user side. Using these tools, packets are captured at the Internet protocol (IP) level. One must note that the cellular traffic is encrypted in layer 2, and hence, the payload of captured traffic is neither intended for our blind prediction/classification, nor accessible for analysis. In the following, we describe a set of  6 features, considered in this study, where all of them are defined over a time interval of $\tau$, as follows: $f_1$: number of  uplink packets; $f_2$: number of downlink packets; $f_3$: size of uplink packets; $f_4$: size of downlink packets; $f_5$: radio of number of uplink to downlink packets; $f_6$: the communication protocol used in the transfer, e.g. TCP or UDP. Based on these features, we define 6 different feature sets (FSs), each containing a subset of features, as follows: FS-1=[1,1,1,1,1,0],  FS-2=[1,0,0,0,1,0], FS-3=[1,0,0,0,0,0],  FS-4=[1,1,0,0,1,0], FS-5=[1,1,0,0,0,0], and FS-6=[1,1,0,0,0,1], where a one (res. zero) at position $i$ of FS-$k$ represents that $f_i$ is present (res. absent) in FS-$k$.


\subsubsection{Experiment setup} \label{sec:setup}
 The  experimental results in the following sections are presented in 3 categories, including i) prediction of number of packet arrivals in the future  time intervals, ii) prediction of burst occurrence  in the future  time intervals, and iii) classification of  applications which are generating the traffic. In the first two categories, we carry out a comprehensive set of Monte Carlo MATLAB simulations \cite{mont},  over the dataset, for different lengths of the training sets, length of future prediction, feature sets used in learning and prediction, and etc.    The notations of  schemes presented in the experiments are as follows: (i) AR(1), which represents predicting the traffic  based on the last observation; (ii) optimized ARIMA, in which the number of lags and coefficients of ARIMA are optimized using a grid search for RMSE minimization in prediction of traffic for the next time interval; (iii) RAF, which combines the results of 50 decision trees for classification, and (iv) LSTM(FS-$x$), in which FS-$x$ for $x\in\{1,\cdots,6\}$ represents the feature set used in the RNN, and the RNN itself consists of one LSTM layer with 100 hidden elements and one fully connected layer. The training of LSTM has been done over 100000 of time intervals of length $\tau$.

\smallskip
\noindent \textbf{Reproducibility}
All experiments could be reproduced using the dataset available at the supporting Github repository \cite{GHAA}.

\subsubsection{Empirical results} \label{sec:results}~Prediction and classification 

\begin{figure}[!htb]
        \centering
                \includegraphics[trim={.6cm 0 .1cm  0},clip,width=3.25 in]{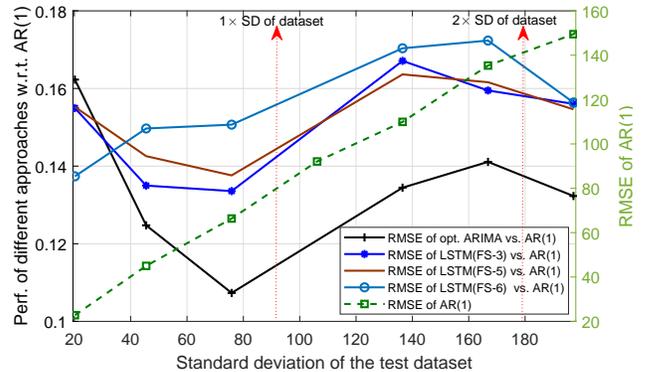}
                \caption{The impacts of traffic type and employed feature sets on the RMSE performance of predictors ($\tau$=10 sec)} 
                \label{comprmse}
\end{figure}

\smallskip
\noindent \textbf{Prediction of traffic arrival}
First, we investigate the performance impacts of traffic type and employed feature sets on the RMSE performance of predictors. Fig. \ref{comprmse} represents the RMSE results for  LSTM predictor with different feature sets, ARIMA with optimized parameters, and  AR(1), when the number of uplink packets in the next time intervals, i.e. 10 seconds, is to be estimated. Towards this end, the right $y$-axis represents the absolute RMSE of AR(1) scheme,  the left $y$-axis represents the relative performance of other schemes versus AR(1), and the $x$-axis represents the standard deviation (SD) of the test dataset. The results are insightful and shed light to the  regions in which ARIMA and LSTM perform favorably, as follows. When the SD of traffic from its average value  is more than 30\% of the long-term SD of the dataset, which is almost the case in the active mode of phone usage by human users, LSTM outperforms the benchmark schemes. On the other hand, when there is only infrequent light background traffic, which is the case on the right-end side of Fig. \ref{comprmse}, ARIMA outperforms the benchmark schemes. When we average the performance over  a 24-days dataset, we observe that  LSTM(FS-6), LSTM(FS-5), LSTM(FS-3), and optimized ARIMA outperform the AR(1) by 16\%, 14.5\%, 14\%, and 12\%, respectively, for $\tau$=10 sec.   Recall that LSTM(FS-6) keeps track of the number of uplink and  downlink packets, as well as  statistics of the communication protocol used by packets in each time interval, while LSTM(FS-5) does not care about the protocol used by packets. The superior performance of LSTM(FS-6) with regards to LSTM(FS-5), as depicted in Fig. \ref{comprmse}, represents that how adding features to the LSTM predictor can further improve the prediction performance in comparison with the linear predictors.

\begin{figure}[!htb]
        \centering
                \includegraphics[trim={.8cm 0 0.8cm 0},clip,width=3 in]{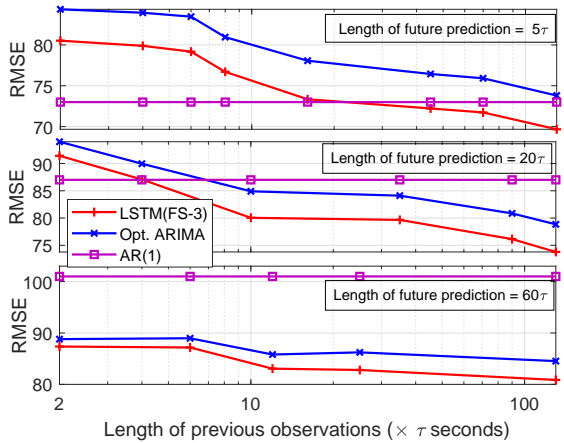}
                \caption{The RMSE performance of LSTSM and ARIMA in short to long-range future traffic prediction ($\tau$=10 sec)} 
                \label{axkol}
\end{figure}

Now, we investigate strengths of different predictors in medium to long-term traffic prediction. Fig. \ref{axkol} represents the RMSE results for 3 different lengths of future predictions, i.e. 50 seconds (top), 200 seconds (middle), and 600 seconds (down).  The $x$-axis represents the length of previous observations, i.e. it represents the number of observations just before the test window, which are available to be used by the trained model. The square-marked curve represents the results for AR(1), i.e. the case in which estimation is made based on the last observation. One observes that for medium-range future prediction, AR(1) outperforms the others when the number of previous observations is less that a threshold value, e.g. approximately 15 observations for $5\tau$-length future observations.  Beyond this threshold value, we observe that LSTM outperforms the AR(1). Furthermore, we observe that this threshold value is dependent on the length of future prediction because in the middle and bottom figures, the LSTM predictor outperforms the others with the threshold value of 4 and 1 previous observations, respectively.  The results further indicate that the optimized ARIMA, which has been optimized for traffic prediction in next interval, loses its performance in longer ranges of future prediction, i.e. it is worse than AR(1) in some circumstances. Finally, as observed in Fig. \ref{comprmse}, the relative performance of LSTM to AR(1) and ARIMA is highly dependent on thr feature set used in training, an hence, the threshold value for LSTM decreases by incorporating further features.

\begin{figure}[!htb]
        \centering
                \includegraphics[trim={.5cm 0 1cm  0},clip,width=.86\columnwidth]{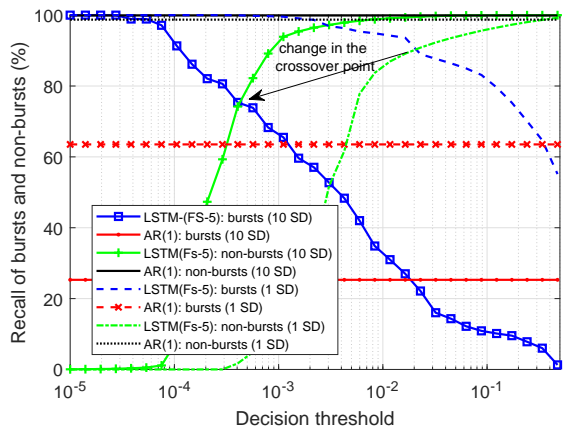}
                \caption{Performance evaluation of prediction of bursts as a function of decision threshold ($\tau$=10 sec)} 
                \label{tradrecalacc}
\end{figure}

\smallskip
\noindent \textbf{Prediction of burst events}
 For the following experiments, we label a subset of time intervals as burst based on the intensity of traffic in each interval, where the intensity could be due to the  number or size  of packets. Then, based on this training dataset, we aim at predicting if a burst will happen in the next time interval(s) or not. As a benchmark to the LSTM predictor, we compare the performance against AR(1), i.e. we estimate a time interval as burst if the previous time interval had been labeled as burst. Fig. \ref{tradrecalacc} represents the recall of bursts and non-bursts for two different burst definitions. The first (second) definition treats the time intervals with more than 90 (900) uplink packet arrivals as burst, when the SD of packet arrivals in the dataset is 90. The LSTM predictor developed in this experiment returns the probability of burst occurrence in the next time interval, based on which, we need to  set a threshold probability value to declare the decision as burst or non-burst.    The $x$-axis of Fig. \ref{tradrecalacc}  represents the decision threshold, which tunes the importance of recall and accuracy of decisions. In this figure, we observe that  the probability of missing a burst is very low in the left side, while the accuracy of decisions is low (it could be inferred from the recall of non-bursts). Furthermore, on the right side of this figure, the probability of missing bursts has been decreased, however, the accuracy of decisions has been increased. The crossover point, where the recall of bursts and non-burst match, could be an interesting point for investigating the prediction performance. In this figure for burst definition of type (1 SD), one  observes that when the decision threshold is 0.02, 91\% of bursts could be predicted, while only 9\% of non-bursts are labeled as burst (false alarm).

\begin{figure}[!htb]
        \centering
                \includegraphics[trim={.6cm 0 .8cm 0},clip,width=0.7\columnwidth]{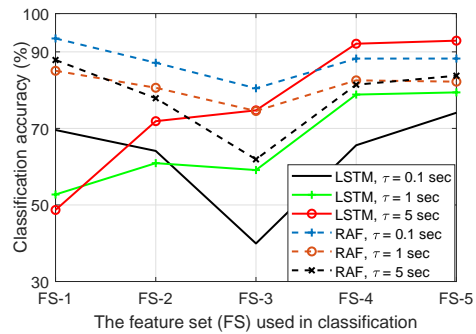}
                \caption{Accuracy of classification by LSTM and RAF as a function of the feature set used in the experiment} 
                \label{accFS}
\end{figure}

\smallskip
\noindent \textbf{Classification of traffic}
Finally, we investigate leveraging machine learning schemes for classification of the application generating the cellular traffic in this subsection. For the classification purpose, a controlled experiment at the  user-side has been carried out in which, 4 popular applications including surfing, video calling, voice calling, and video streaming have been used by the user. Fig. \ref{accFS} represents the  accuracy of classification for different feature sets used in classifiers, i.e. LSTM and RAF.  For the case of LSTM, one observes that the LSTM(FS-5) outperform the others significantly in the accuracy  of classification, and the accuracy increase in $\tau$. On the other hand, one observes that the RAF scheme achieves the best performance for FS-1, i.e., when it has full access to all features, and it performance decreases by an increase in $\tau$. The reason for the former difference in behavior (best accuracy in FS-1 or FS-5) consists in the fact that LSTM(FS-1) suffers from over-fitting, while the RAF can compensate this problem by averaging over many decision trees. Then, if a few number of features are available, LSTM performs preferably, and vice versa.  On the other hand, as $\tau$ increases, due to practical problems with short $\tau$ values, the ambiguity in making decision for RAF increases, while the LSTM can make a better decision thanks to its sophisticated design. 


%

 \begin{table}[!htb]
\centering \caption{Parameters for performance evaluation}\label{sim3}
\begin{tabular}{p{3.2 cm} p{4.7 cm}}\\
\toprule[0.5mm]
{\it Parameters }&{\it Value}\\
\midrule[0.5mm]
Service area& Cell of radius 500m, BS at center\\
Average service rate& 45 Mbps\\ 
BS transmit power& Adaptive to user channel to fulfill on-average 45 Mbps, Max: 40 W \\
Type-1 traffic&SPP($0.2$,$20$,$10$,$1$), Size: 3Mb\\
Type-2 traffic&PP($2$), Size: 2Mb\\ 
Number of  users & 5 of type-1; 3 of type-2\\
Resource management& Round robin scheduling\\
 \bottomrule[0.5mm]
\end{tabular}
\end{table}

 \begin{figure}[!htb]
        \centering
                \includegraphics[trim={.2cm 0 .1cm 0},clip,width=0.85\columnwidth]{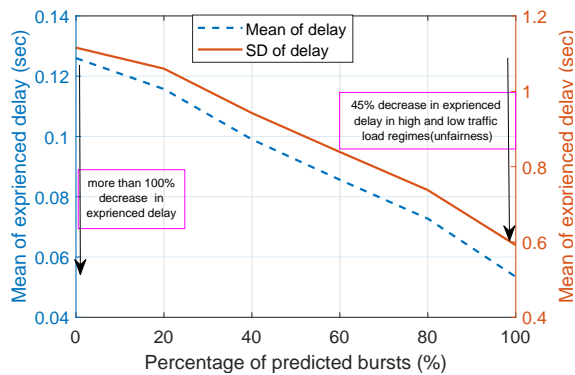}
                \caption{The impact of burst prediction and predictive buffering on the QoS in terms of service delay for type-2 traffic} 
                \label{del}
\end{figure} 


  \subsection{Traffic-aware resource management}
In this section, we conduct a case study to investigate the performance impact of prediction of burst arrivals on the experienced delay in communications. Consider a service area, with one BS at the center, serving two types of bursty and non-bursty downlink service requests, modeled by Switched Poisson Process (SPP) and Poisson Process (PP), respectively \cite{BSS}. The type-1 and type-2 traffic models aim at representing  simplified models of  surfing/on-demand file downloading and streaming applications, respectively, where the later is more sensitive to delay in communications. The parameters for SPP in Table \ref{sim3} represent traffic arrival in light and heavy traffic periods, and the average lengths of  light and heavy traffic periods, respectively. For the PP model, the parameter represents the traffic arrival rate. In our Matlab simulator, once a burst in type-1 traffic is predicted, the BS starts filling the buffer of users which are served by type-2 traffic, and hence, in order to prevents QoS degradation for type-2 traffic at the time of arrival of burst for type-1 traffic. 
Fig. \ref{del} represents the impact of burst prediction in   type-1 traffic on the experienced delay by users requesting type-2 traffic. The $x$-axis in this figure represents the percentage of predicted burst, as per the results of 
Fig.~\ref{tradrecalacc}.  One observes that the expected service delay in for type-2 traffic could be significantly decreased by predicting burst in type-1 traffic. One further observes that burst prediction also  significantly decreases the standard deviation of delay, i.e. the severe impact of occurrence of bursts, are compensated. These promising results motivate jointly formulating the radio resource allocation and user traffic prediction, and driving probabilistic schedulers, which are skipped here.

\section{Conclusions} \label{sec:conclusions}
In this work, the feasibility of per-user traffic prediction for cellular networks has been  investigated. Towards this end, a framework for cellular traffic prediction has been introduced, which leverages statistical/machine learning tools for traffic classification and prediction. A comprehensive comparative analysis of traffic prediction based on statistical  and deep learning has been carried out, under different traffic circumstances and design parameter selections.  The LSTM model, when additional traffic statistic features are accessible or there is access to a set of previous observations, exhibited demonstrable improvement over the optimized ARIMA model for  short to long-term future predictions. The impact of number of previous observations, length of future prediction, type of features in use, and type of application(s) generating the traffic on the accuracy of predictions have been investigated, and it has been shown, and the circumstances in which statistical, rule-based and deep learning approaches perform favorably have been highlighted. 
Furthermore,  usefullness of the developed learning tools for classification of cellular traffic has been investigated, where the results represent high sensitivity of accuracy and recall of classification to the feature set in use. Our simulations, for a radio resource management problem,  have shown  a considerable decrease in experienced delay, when the decision making module is augmented by burst traffic arrival estimation.

\ifCLASSOPTIONcaptionsoff
  \newpage
\fi

\bibliographystyle{IEEEtran}
\bibliography{paperbib}

\end{document}